# Solvation Structures and Ion Dynamics of CaCl$_2$ Aqueous Electrolytes Using Metadynamics and Machine Learning Molecular Dynamics Simulations

Zhou Yu, Lei Cheng*

Materials Science Division and Joint Center for Energy Storage Research, Argonne National Laboratory, Lemont, Illinois 60439, United States

*Correspondence: Lei Cheng: chengl@ornl.gov

**Keywords:** molecular dynamics simulations, metadynamics, machine learning potentials, solvation structures, ion dynamics, CaCl$_2$ aqueous electrolytes

**Abstract:** The solvation structures and ion dynamics of CaCl$_2$ aqueous electrolytes have been investigated using *ab initio* molecular dynamics simulations and molecular dynamics simulations with deep learning potentials. We found multiple solvation structures around the Ca$^{2+}$ ion, including fully hydrated single Ca$^{2+}$ ion, Ca-Cl contact ion pair, and Ca-2Cl bridged ion pair, could coexist. The ion-pairing condition plays an important role in the translational and orientational distribution of water molecules in the solvation shell. And the local ordering introduced by the Ca$^{2+}$ ion can extend to the second solvation shell. Furthermore, we found the lifetime of water molecules in the solvation shell is sensitive to the ion-pairing conditions. The self-diffusivities of ions and water molecules, as calculated in molecular dynamics simulations with deep learning potentials, are in good agreement with experimental measurements. Finally, we elucidate the transition of Ca$^{2+}$ ion dynamics between different regimes by analyzing angle probability distribution histograms and van Hove correlation function.

## 1. Introduction

Aqueous electrolytes are ubiquitous components in chemistry, energy, biology, and environmental sciences.[1,2] Over the years, numerous noteworthy studies have emerged aimed at comprehending the solvation structure of ions, dynamics of ions and water, and macroscopic thermodynamics, all of which play crucial roles in the design and application of aqueous electrolytes. For example, investigations employing techniques such as dielectric relaxation and femtosecond infrared spectroscopy have revealed that monovalent cations and anions can exhibit a cooperative effect on slowing down the reorientation dynamics of the water molecules in hydration shells.[3,4] Far-infrared (terahertz) and X-ray absorption studies also provide direct evidence of the formation of contact ion pair.[5] Nevertheless, the solvation structures captured through diverse scattering and electroscopic techniques remain inconclusive.[6-8] Additionally, debates persist regarding the range of ions' effect on the water. X-ray absorption suggests a local effect, limited to one hydration shell,[9] and dielectric relaxation support a long-range effect, extending to several hydration shells.[3] For a more detailed understanding, interested readers are encouraged to refer to several comprehensive reviews discussing the experimental investigations on the structure and dynamics of hydration ions.[5,10-12]

From a theoretical standpoint, two computational methodologies have been extensively employed to investigate the physiochemical properties of aqueous electrolytes: hybrid cluster/continuum quantum chemistry calculations and molecular dynamics (MD) simulations. The former is typically employed to explore the solvation structure and thermodynamics (e.g., solvation energy, redox potential, etc.).[13,14] This approach heavily relies on ion charge localization and the relevance of the infinite dilution limit.[15] MD simulations, encompassing classical and *ab initio* methods, can provide ensemble-averaged solvation structure, dynamics, and thermodynamics.[16,17] Classical MD (CMD), in comparison to *ab initio* MD (AIMD), offers a longer time scale (~100 ns) and larger length scale (~10 nm). Therefore, CMD has been broadly applied in complex systems composing of intricate and bulky ions and solvent molecules.[18,19] However, the accuracy of CMD highly depends on the reliability of the potential/force field. Some recent studies emphasize the importance of considering the electronic diversity of water molecules[20] and charge transfer between an ion and its hydration shell[21] to accurately reproduce the dynamics in aqueous electrolytes. AIMD simulations, which utilize forces derived from the ground state of electrons, can be used to capture the accurate structure, dynamics, and thermodynamics. However, some transitions between metastable solvation structures are hindered kinetically and cannot be observed within the short-time period currently accessible to AIMD (~100 ps). Therefore, AIMD simulations are primarily applied in some simple scenarios, such as single-ion systems (a single cation/anion mediated in solvents) and monovalent ion pair systems.[15, 21, 22] Recent advancements in machine learning force fields have partially resolved the accuracy versus efficiency dilemma.[23-26] A deep learning potential can be trained using a large database comprising corresponding atomic coordinates, forces, and energies calculated from AIMD simulations or density functional theory (DFT) calculations. Recent studies have demonstrated that MD simulations employing deep learning potentials can accurately reproduce the structural and dynamic properties of several condensed phase systems obtained from AIMD simulations while significantly reducing computational costs by a factor of $10^4 \sim 10^5$.[27, 28]

In this work, we employ a combination of metadynamics AIMD and machine learning MD simulations to investigate the solvation structure and ion dynamics in a representative divalent aqueous electrolyte, specifically, $CaCl_2$. To the best of our knowledge, this is the first investigation to systematically examine solvation and dynamics in a multivalent aqueous solution using the combination of metadynamics and machine learning techniques. Our findings reveal the coexistence of multiple metastable solvation structures within the solution. Furthermore, we highlight the crucial role of ion pairing in shaping the translational and orientational distribution of water molecules within the solvation shell, as well as the lifetime of water molecules in that shell. Additionally, we demonstrate that MD simulations employing a deep learning potential are highly effective in capturing ion dynamics with accuracy comparable to AIMD simulations.

## 2. Computational Methods

### 2.1. Simulation systems

The metadynamics AIMD simulations[29] were performed to explore the free energy surface (FES) as a function of the coordination environment of $Ca^{2+}$ ion in the $CaCl_2$ aqueous solution. 1 $CaCl_2$ and 64 water molecules were randomly packed in a cubic simulation box with a side length of 12.51 Å using the PACKMOL code.[30] The concentration is 0.848 M (mol/L) and the density is 1.073 g/ml, which is consistent with the experiments.[31] The initial configuration was then relaxed using classical MD simulations with generic OPLS-aa force fields[32] under the NVT ensemble for 1 ns. Then, the AIMD metadynamics simulations were performed for 100 ps. The normal AIMD without biased potential and machine learning MD simulations[23] were performed starting from three local metastable solvation structures of $Ca^{2+}$ ion, which are fully hydrated single $Ca^{2+}$ ion, Ca-Cl contact ion pair, and Ca-2Cl bridged ion pair, denoted as FHS, CIP, and BIP system, respectively. The dimension and composition of these three simulations are consistent with metadynamics AIMD simulations. Three AIMD and MD simulations were performed for 25 ps and 1 ns, respectively.

## 2.2. Simulation methods

The AIMD simulations were carried out using the Quickstep module implemented in the CP2K package[33] under the NVT ensemble. Core electrons were described by norm-conservative Goedecker-Teter-Hutter (GTH) pseudopotentials.[34] The valence electrons were described at the DFT level using the revPBE parametrization functional[35] and an atom-centered Gaussian double-zeta basis set augmented with a set of d-type or p-type polarization functions (DZVP-MOLOPT-SR-GTH).[36, 37] Several prior studies have found the revPBE functional has great performance in reproducing the dynamics properties in water-based systems.[38-40] The DZVP-MOLOPT-SR-GTH has also been broadly applied in many liquid electrolytes.[41, 42] The energy grid cutoff was set as 400 Ry, which was determined through the energy converge test (shown in Figure S1). DFTD3 empirical Grimme correction is considered for the long-range dispersive forces.[43] The MD simulation with the deep learning potential was performed using LAMMPS[44] under the NVT ensemble. Nose-Hoover thermostat was used to stabilize the temperature at 303 K.[45, 46] The simulation box is periodic, and the time step is 0.5 fs for both AIMD and MD.

Two collective variables, the coordination number (CN) of $Cl^-$ ions and oxygen atoms in water molecules around the $Ca^{2+}$ ion, were used in metadynamics AIMD simulations. Small repulsive Gaussian hills with a height of ~0.054 eV (0.002 Hartree) and a width of 0.2 CN were added at a frequency of once per every 100 time steps for these two collective variables. The low Gaussian hill, which is ~1/60 of the depth of the deepest valley in the free energy surface (FES), and the narrow width make sure we can capture precise FES as a function of the solvation structure. Meanwhile, the system can relax between two biased potentials, and the energy drift is less than 1 meV atom$^{-1}$ ps$^{-1}$. The representative solvation structures (e.g., FHS, CIP, and BIP) in the FES maintained in the last 10 ps of the simulations (see Figure S2), which proves the AIMD metadynamics simulations were converged. The CN values collected in the metadynamics simulation were defined as $CN = \sum_{i=1}^{N} \frac{1-(r_i/r_o)^p}{1-(r_i/r_o)^q}$,[47] where $p$=12, $q$=24, $r_o$ is the cutoff distance (i.e., 3.4 Å), which corresponds to the first valley in the radial distribution function (RDF) from $Ca^{2+}$ to $Cl^-$ ion and O in water within the AIMD simulations (see Figure S3a and Figure 2a); and $r_i$ is the distance from $Ca^{2+}$ to the $i^{th}$ $Cl^-$ ion or O atom.

### 2.3. Deep learning potential

The deep learning potential was constructed following the method proposed by Zhang et al.[23, 48] The potential energy of a system can be decomposed into a sum of atomic energy contributions, and the atomic energy is fully determined by the coordinates of this atom and its neighboring atoms within a smooth cutoff radius. Explicitly, the deep learning potential was constructed in two steps, including preparing data and training the model. First, a large database composed of 50000 data sets has been prepared. One data set includes the atomic coordinates, corresponding atomic forces, and static energy. Twenty thousand coordinates were extracted from 100 ps metadynamics simulations at a time interval of 5 fs. The static force and energy were calculated on each coordinate with the same DFT method as the AIMD simulation using the Quickstep module in the CP2K package.[33] The other 30000 data sets were from the last 20 ps of the three AIMD simulations on the local solvation structures at a time interval of 2 fs. Then, the whole database was passed to Tensorflow[49] for deep neural network training using the DeepMD-kit package.[48] At the end of the training, the root mean square energy and force errors over the whole testing set are $2.5\times10^{-4}$ eV and $6.6\times10^{-2}$ eV/Å shown in Figure S4. We refer the reader to the recent work from Wang and E's group for the detailed theoretical discussion.[23, 48, 50, 51]

### 3. Results and Discussion

### 3.1 Solvation Structure

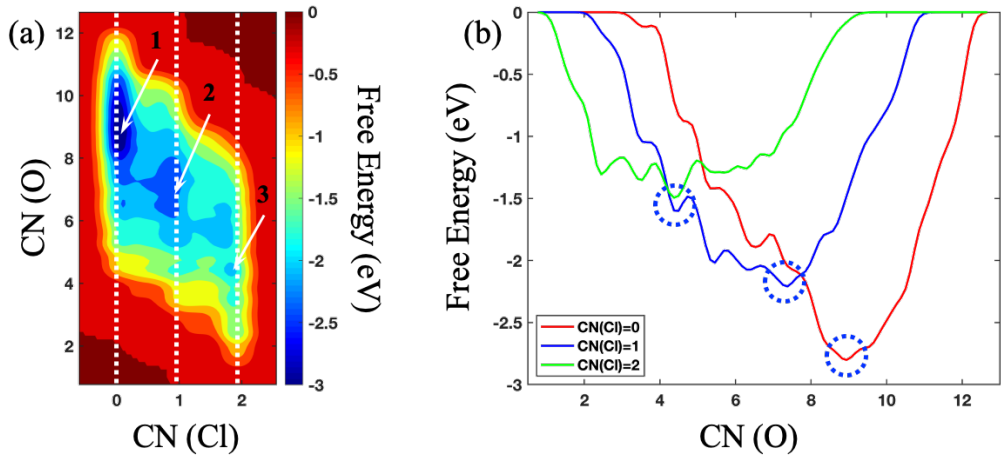

Figure 1. (a) Free energy surface contour plots of the CaCl$_2$ aqueous solution as a function of CN of Cl ($x$-axis) and O ($y$-axis) around the Ca$^{2+}$ ion from metadynamics AIMD simulations. (b) Free energy profiles as a function of CN of O around the Ca$^{2+}$ ion with 0, 1, and 2 CN of Cl$^-$ ions, corresponding to three white dotted lines in (a).

We employed metadynamics AIMD sampling for the CaCl$_2$ aqueous solution to explore the solvation environment composing of O in water and Cl$^-$ ion around Ca$^{2+}$ ion. Our analysis of the free energy landscapes revealed the presence of distinct local free energy minima, designated as 1, 2, and 3 in Figure 1a. These minima correspond to specific configurations: the fully hydrated single Ca$^{2+}$ ion (FHS), Ca-Cl contact ion pair (CIP), and Ca-2Cl bridged ion pair (BIP). Although the FHS configuration has the lowest free energy, the mild free energy barrier (< 1 eV) for Ca$^{2+}$ to exchange through these free energy minima implies that both ion pair and fully dissociated configurations likely coexist in ~0.85 M CaCl$_2$ aqueous solution. To gain a clearer understanding of the free energy profiles along these configurations, we analyzed the data along the three white dotted lines in Figure 1a, as illustrated in Figure 1b. For the FHS configuration, the local free energy minima locate around 9 CN of O, which is consistent with the experimental hydration number through vapor pressure measurements.[52] When one Cl$^-$ ion involves in the first solvation shell around the Ca$^{2+}$ ion, the local free energy minima is with ~7 CN of O. Meanwhile, a plateau appears in the range of 5 to 7 CN of O, which is consistent with the observation in a recent work[53] that multiple stable CN of O (i.e., 5 and 6) in the solvation shell of CIP can coexist. When two Cl$^-$ ions are in the first solvation shell of the Ca$^{2+}$ ion, a wider plateau exists from 2 to 6 CN of O. The energy wells and profiles of both CIP and BIP configurations are shallower and more dispersed, indicating less-defined solvation structures. Conversely, the FHS configuration exhibits a relatively deep and narrow energy well, suggesting a pronounced preference for a specific solvent coordination configuration. These findings underscore the diversity of solvation structures present in CaCl$_2$ aqueous solution and corroborate the theory of multiple stable solvation structures.[15]

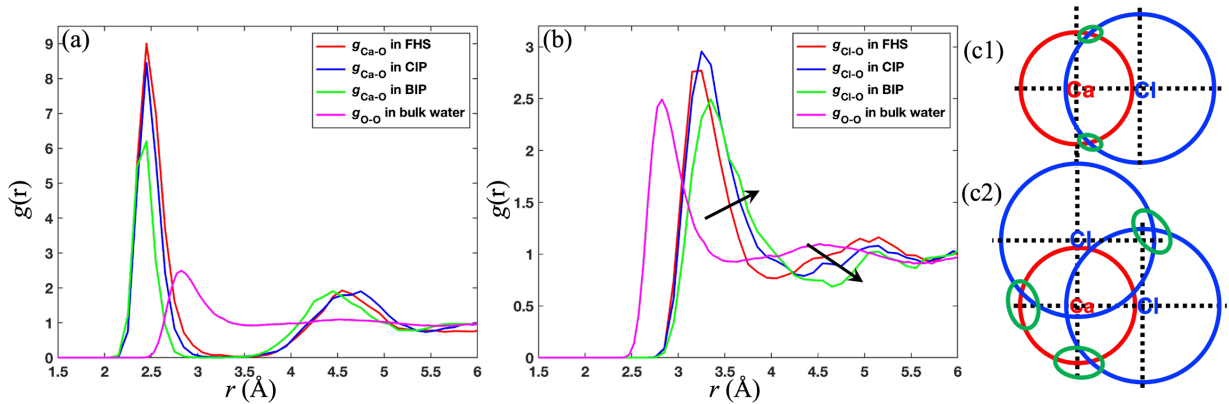

Figure 2. RDF from (a) Ca to O ($g_{Ca-O}$) and (b) Cl to O ($g_{Cl-O}$) in FHS, CIP, and BIP system. Magenta curves in (a-b) are the RDF among O atoms in bulk water ($g_{O-O}$). Scheme of the interference on the distribution of O around Ca and Cl ions in (c1) CIP system and (c2) BIP system. The red and blue circle denote the first peak of O around Ca and Cl in FHS system, respectively. The green circle represents the interference region.

Following 25 ps of normal AIMD simulations without biased potentials, we found that the initial solvation structures (FHS, CIP, and BIP) remained stable across all three systems. This observation serves as a cautionary note for researchers regarding the selection of initial configurations in AIMD simulations. Within the limited simulation time, solvation structures can easily become trapped in metastable conditions. Therefore, careful consideration should be given to the initial setup of simulations. Figure 2a reveals that the first peak of $g_{Ca-O}$ locates at ~2.45 Å in all three cases and becomes lower and narrower when there are more Cl$^-$ ions in the first solvation shell around the Ca$^{2+}$ ion. A distinct second solvation shell can be observed at ~4.6 Å from the center Ca$^{2+}$ ion. Slight differences on the second peak may originate from the free Cl$^-$ ion, which also affects the water distribution. A recent study found the local ordering introduced by the Na$^+$ ion in the first solvation shell does not extend further into the liquid since a highly consistent distribution occurs between $g_{O-O}$ and $g_{Na-O}$ beyond the first peak.[22] However, the local ordering introduced by Ca$^{2+}$ ion can extend to a much longer distance, since $g_{O-O}$ and $g_{Ca-O}$ exhibit unique distribution at the second solvation shell. Similar long-distance distribution has also been observed around Mg$^{2+}$ ions in aqueous solution in our recent work.[54] This observation highlights the multivalent cation has the capacity to lead a longer range ordering compared to the monovalent cation.

The $g_{Cl-O}$ shifts to a larger distance (see the arrows in Figure 2b) beyond the first peak at ~3.3 Å when there are more Cl⁻ ions in the first solvation shell around the $Ca^{2+}$ ion. This can be understood through the interference between the water distribution around $Ca^{2+}$ and Cl⁻ ion shown in Figure 2c. When $Ca^{2+}$ and Cl⁻ ion form the CIP configuration, the hydration shell will overlap to form an interference region shown in green circles in Figure 2c. Since the first peak in $g_{Ca-O}$ is much higher than that in $g_{Cl-O}$, more free energy is needed to be overcome for a water molecule escaping from the $Ca^{2+}$ ion's hydration shell than Cl⁻ ion's.[55] In other words, the first hydration shell around the $Ca^{2+}$ ion is more rigid than that around the Cl⁻ ion. Consequently, the interference will have a more significant effect on the distribution of water around the Cl⁻ ion. For instance, in Figure 2c1, the positive interference within the green circle between the first hydration shell of the $Ca^{2+}$ ion and the Cl⁻ ion results in a slight compression of the surrounding water molecules. In the BIP system, there is a positive interference between the $Ca^{2+}$ ion and the two attached Cl⁻ ions. Additionally, given that the preferred ∠ClCaCl is approximately 90° (see Figure S3b), another interference occurs between two Cl⁻ ions in the BIP system. Consequently, $g_{Cl-O}$ can extend to a longer distance. This interference theory has already been successfully used to explain colloid and polymer interaction in inorganic salts and ionic liquids.[56, 57]

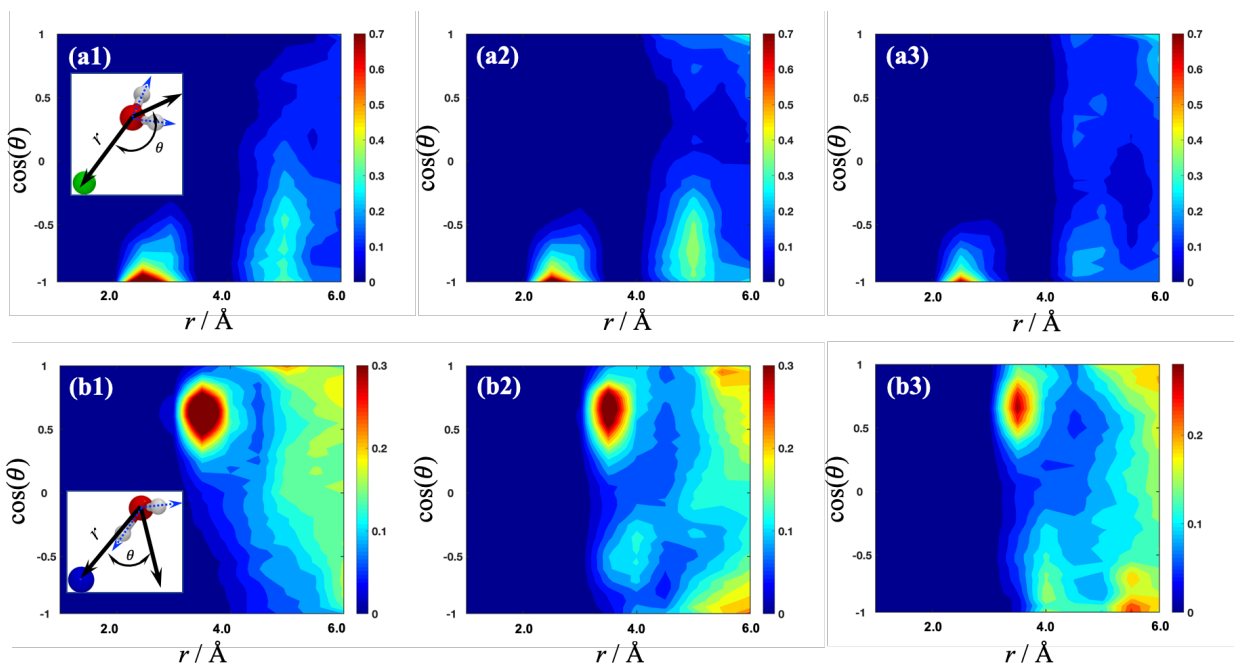

Figure 3. 2D Angular-radial distribution functions of ∠XO$_c$H$_2$ and XO$_c$. (X is $Ca^{2+}$ ion in (a) and Cl⁻ ion in (b), O$_c$H$_2$ represents the bisector of the water molecules, and O$_c$ represents O in water molecules). The three

columns represent FHS, CIP, and BIP case from left to right. The definition of angle and distance is also shown in the set of (a1) and (b1) for clarify. The color bar denotes the time averaged frequency.

The 2D angular-radial distribution function[58] provides information on the preferred orientation of water dipoles near $Ca^{2+}$ and $Cl^-$ ion in different solvation shells. In our analysis, we calculated the angular distribution between $O_cX$ and the bisector of the water molecules $H_2O_c$, as well as the radial distribution between X and $O_c$. Here, X denotes an ionic center, and $O_c$ represents the oxygen atom within a water molecule. From Figure 3(a1-a3), we can see the angular distribution of water molecules in the first solvation shell ($r$=2.0~3.0 Å) around the $Ca^{2+}$ ion is very similar, which mainly locates around $\cos(\theta)$ = -1 ($\theta$=180°). The influence of the $Ca^{2+}$ ion on neighboring water molecules resembles that of a hydrogen bond donor in bulk water. For the water molecules in the second solvation shell ($r$=4.0~6.0 Å) around the $Ca^{2+}$ ion, the angular distribution becomes broader. Moving from FHS to CIP and subsequently to BIP, we observe a decrease in the frequency of large angle distribution (-1.0<$\cos(\theta)$<0), while new features emerge in the distribution at small angles (0<$\cos(\theta)$<1.0). These changes primarily arise from the effects of the attached $Cl^-$ ion on the surrounding water molecules.

From Figure 3(b1-b3), we observe distinct features in the angular distribution of water molecules in the first solvation shell ($r$=3.0~5.0 Å). For instance, the peak appears in the range of $\cos(\theta)$ = 1.0 to 0.5 in the FHS system. This corresponds to the value of $\theta$ from 0 to 60°, indicating that the influence of the single $Cl^-$ ion on the neighboring water molecules resembles that of a hydrogen bond acceptor in bulk water. This peak becomes less prominent, and another board preferential region appears around $\cos(\theta)$ =0 to -1.0 (see Figure 3b2-b3) when $Cl^-$ ion approaches the $Ca^{2+}$ ion to form an ion pair. This originates from the significant effects of the $Ca^{2+}$ ion on the angular distribution of the water molecules. Regarding the angular distribution of water molecules within the second solvation shell around the $Cl^-$ ion ($r$>5.0 Å), it becomes smeared due to the combined effects of preferred orientation around the $Ca^{2+}$ ion and hydrogen bonding. Consequently, no clear preference for the tilt angle is observed for these water molecules.

## 3.2 Dynamics of ions and solvation shell

To overcome the time-consuming nature of AIMD simulations, MD simulations with the deep learning potential were conducted to investigate the dynamics in the $CaCl_2$ aqueous solution. The reliability of the MD simulations with deep learning potential has been thoroughly validated against AIMD simulations, encompassing various structural and dynamic parameters. Further details can be found in the supporting information section 3. To characterize the influence of ion pairing on the dynamics of water molecules, we calculate the residence correlation function of water molecules within the first solvation shell surrounding the $Ca^{2+}$ and $Cl^-$ ions. The frequency of water exchange around ions is widely acknowledged as a crucial factor for the reactions involving these ions in aqueous solutions.[59, 60] The residence correlation function is defined as $\langle c(0)c(t) \rangle$. $c(t)$ is equal to 1.0 if the water molecule within the solvation shell at $t = 0$ continuously resides in this shell until time $t$. Otherwise, $c(t)$ equals 0. By examining the time evolution of the residence correlation function illustrated in Figure 4, we observe that it becomes more challenging for water molecules to escape from the first solvation shell around the $Ca^{2+}$ ion as more $Cl^-$ ions become involved in the solvation shell. Meanwhile, water molecules can escape more readily from the single $Cl^-$ ion compared to the $Cl^-$ ions involved in ion pairing.

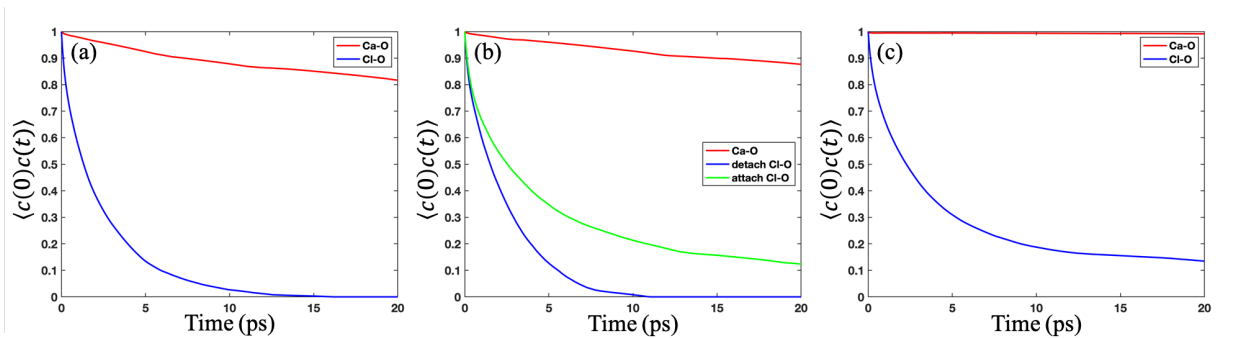

Figure 4. Residence correlation function for water in the first solvation shell around $Ca^{2+}$ and $Cl^-$ ion in (a) FHS, (b) CIP, and (c) BIP system. The cutoff distance for water molecules in the first solvation shell corresponds to the first valley in $g_{Ca-O}$ and $g_{Cl-O}$ shown in Figure 2a-b.

By examining the trajectories of machine learning MD simulations in three distinct systems, we found both CIP and BIP systems transform to the FHS condition after tens of picoseconds. This transformation occurs because the FHS condition exhibits a lower free energy state compared to the other structures shown in Figure 1. While the CIP and BIP systems can only maintain their initial configurations for a brief period, our subsequent focus primarily centers on the dynamic properties observed in the FHS system. First, we calculated the time evolution of mean square displacement (MSD) of $Ca^{2+}$, $Cl^-$, and water in the FHS utilizing a 500 ps equilibrated trajectory. Diffusion coefficients were then determined by analyzing the MSDs within the diffusive regimes, characterized by a slope of 1. The diffusion coefficients of $Ca^{2+}$, $Cl^-$, and water in the MD simulation is $0.52 \times 10^9$, $1.87 \times 10^9$, and $2.28 \times 10^9$ m$^2$/s. Compared to experimental diffusion coefficients of $Ca^{2+}$ ($0.66 \times 10^9$ m$^2$/s), $Cl^-$ ($1.59 \times 10^9$ m$^2$/s), and water ($1.89 \times 10^9$ m$^2$/s),[61] the differences are less than 22%. Given the possibility of the presence of a small fraction of ion pair conditions in the solution and the temperature disparity between the experimental setup (296.15~298.15 K) and our simulations (303.15 K), these discrepancies in the diffusion coefficients are considered negligible. This remarkable agreement underscores the effectiveness of MD simulations employing deep learning potential as a formidable tool for studying ion dynamics over extended trajectories, surpassing the traditional time scale limitations of AIMD (~100 ps).

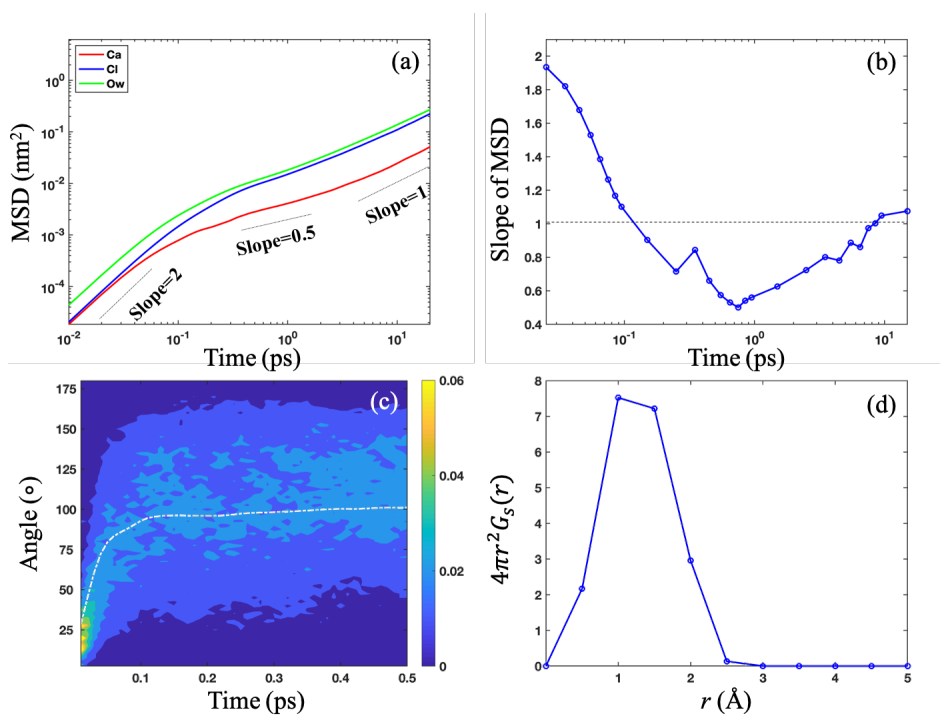

Figure 5. (a) MSD curves of $Ca^{2+}$, $Cl^-$, and water in the FHS system. (b) Slope of MSD of the $Ca^{2+}$ ion. (c) Histograms of the angle probability distribution as a function of the time interval Δ for 0-0.5 ps of Ca ion in the FHS system. (d) Self-part van Hove correlation function $4\pi r^2 G_s(r)$ for the Ca ion in the FHS system.

Furthermore, it is crucial to acknowledge the significant influence of the solvation shell on ion dynamics. Initially, the motion of the ion remains unhindered by its solvation shell, resulting in a ballistic regime of ion diffusion (characterized by a slope of 2 in MSD). Subsequently, the solvation shell begins to impede the ion's motion, causing the solvation shell to remain relatively stationary while the solvated ion continues to move (yielding a slope of 0.5 in MSD). Finally, as the solvation shell either moves along with the solvated ion or undergoes rupture due to the cleavage caused by the solvated ion, the ion diffusion enters the diffusive regime (manifested by a slope of 1 in MSD).[62] The distinct progression through these three stages is aptly illustrated by the MSD plot of the $Ca^{2+}$ ion depicted in Figure 5b. Notably, the transition between ballistic and subdiffusive regimes occurs at ~0.1 ps. At ~10 ps, the diffusion of the $Ca^{2+}$ ion experiences the transition between subdiffusive and diffusive regimes.

In order to gain further insights into the transition dynamics of the $Ca^{2+}$ ion within different regimes, we conducted additional calculations. Specifically, we investigated the angle probability distribution histograms and self-part van Hove correlation function $4\pi r^2 G_s(r)$ for the $Ca^{2+}$ ion in the FHS system. To analyze the angle probability distribution, we calculated the ensemble average of the angle between two vectors as a function of the time interval. These vectors, denoted as $V_1(t)$ and $V_2(t)$, were obtained by evaluating $X(t+\Delta)-X(t)$ and $X(t+2\Delta)-X(t+\Delta)$, respectively, where $X(t)$ represents the coordinate of the $Ca^{2+}$ ion at time $t$, and $\Delta$ represents the time interval. From Figure 5c, we can see that, initially, a prominent angle distribution emerges at a small angle (~25°). This finding suggests that the ion tends to move in a similar direction as the previous time step, with its motion mildly impeded by the solvation shell. The motion of the $Ca^{2+}$ ion is considered a forward ballistic motion. As the time interval increases, the angle distribution gradually expands, indicating a progressive hindrance of the $Ca^{2+}$ ion's motion by the solvation shell. After approximately 0.1 ps, the angle distribution stabilizes around 100°, indicating that the ion's movement becomes obstructed by other atoms, causing a change in direction. Intriguingly, this time interval aligns with the transition observed between the ballistic and subdiffusion regimes, as depicted in the MSD curve of Figure 5ab.

Additionally, we examined the self-part of the van Hove correlation function, denoted as $G_s(r,t)$, for the $Ca^{2+}$ ion. This function measures the probability that a particle has moved to a distance $r$ away from its initial position at time $t$ = 0. Computed as $G_s(r,t) = \frac{1}{N}\sum_{i=1}^{N}\langle\delta[\mathbf{r}+\mathbf{r}_i(0)-\mathbf{r}_i(t)]\rangle$, where $\delta(\cdot)$ represents the delta function, $\mathbf{r}_i$ denotes the position of ion $i$, this function provides valuable insights into the ion's displacement behavior. Figure 5d illustrates the resulting $4\pi r^2 G_s(r,t)$ for the $Ca^{2+}$ ions at t = 10 ps. Notably, the graph reveals a high probability of the $Ca^{2+}$ ion moving within a range of 0.1~0.15 nm. In fact, the ion can even traverse distances up to 0.25 nm from its initial position. Interestingly, this range corresponds to the first peak observed in the $g_{Ca-O}$ plot shown in Figure 2a, suggesting that the transition from the subdiffusive to the diffusive regime occurs when the motion of the $Ca^{2+}$ ion begins to approach the location of its original solvation shell. Overall, by examining both the angle probability distribution and the self-part van Hove correlation function, we have gained valuable insights into the transition dynamics of the $Ca^{2+}$ ion, shedding light on its behavior within different regimes.

4. **Conclusions**

Using a combination of AIMD and MD simulations with deep learning potentials, we conducted an extensive investigation into the solvation structure and ion dynamics within $CaCl_2$ aqueous electrolytes. Our findings shed light on key aspects of this complex system. Initially, our analysis revealed the coexistence of multiple metastable solvation structures around the $Ca^{2+}$ ion in approximately 0.85 M $CaCl_2$ aqueous solution. Interestingly, we observed that the translational and orientational distribution of water molecules within the solvation shell surrounding the $Ca^{2+}$ ion exhibited sensitivity to the ion pairing conditions, namely FHS, CIP, and BIP. Moreover, our study elucidated the crucial role played by ion pairing conditions in determining the lifetime of water molecules within the solvation shell. Specifically, we discovered that the presence of more $Cl^-$ ions within the solvation shell rendered it more challenging for water molecules to escape the immediate vicinity of the $Ca^{2+}$ ion. Importantly, our investigation underscored the effectiveness of machine learning MD simulations as a powerful approach for studying ion dynamics in a highly efficient manner. Notably, the self-diffusivity of ions and water molecules calculated through our MD simulations yielded results comparable to experimental measurements, highlighting the fidelity of our approach. To gain a comprehensive understanding of the transition dynamics of the $Ca^{2+}$ ion, we employed angle probability distribution histograms and van Hove correlation functions. Through these analyses, we made notable observations. The transition from ballistic to subdiffusive motion occurred when the forward motion of the ion became impeded, while the transition from subdiffusive to diffusive motion manifested when the ion's movement approached the location of its original solvation shell. In short, our study provides valuable insights into the solvation structure and ion dynamics within $CaCl_2$ aqueous electrolytes. By combining AIMD and MD simulations with deep learning potentials, we have unraveled the coexistence of solvation structures, the influence of ion pairing conditions on water molecule lifetimes, and the transitions in $Ca^{2+}$ ion dynamics. Our findings highlight the efficacy of machine learning MD simulations as a robust tool for investigating complex ion dynamics.

**Conflict of Interest**

The authors declare that the research was conducted in the absence of any commercial or financial relationships that could be construed as a potential conflict of interest.

**Author Contributions**

LC led the scientific research and ZY wrote the full paper. LC is the corresponding author of the article.

**Funding**

This research was supported by the Joint Center for Energy Storage Research (JCESR), a U.S. Department of Energy, Energy Innovation Hub. The submitted manuscript has been created by UChicago Argonne, LLC, Operator of Argonne National Laboratory ("Argonne"). Argonne, a U.S. Department of Energy Office of Science laboratory, is operated under contract no. DE-AC02-06CH11357.

**Acknowledgments**

We gratefully acknowledge the computing resources provided on Bebop, a high-performance computing cluster operated by the Laboratory Computing Resource Center at Argonne National Laboratory.

**Supporting Information**

The supporting information includes energy cutoff in AIMD simulations; the evolution of free energy surface (FES); $g_{Ca-Cl}$ in CIP and BIP system; angle distribution of ∠ClCaCl in the BIP system; and validation on MD simulations with the deep learning potential.

**Supporting information**

**Solvation Structures and Ion Dynamics of CaCl$_2$ Aqueous Electrolytes Using Metadynamics and Machine Learning Molecular Dynamics Simulations**

Zhou Yu, Lei Cheng*

Materials Science Division and Joint Center for Energy Storage Research, Argonne National Laboratory, Lemont, Illinois 60439, United States

## 1. Energy cutoff in AIMD simulations

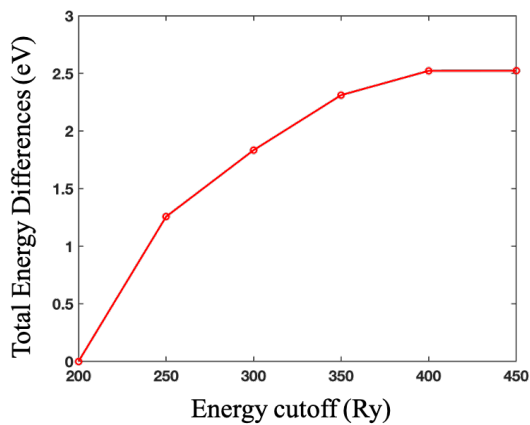

Figure S1. Total energy changes as a function of the energy cutoff

The total energy is converged when the energy cutoff is 400 Ry. The energy change is ~0.002 meV/atom when the cutoff energy changes from 400 to 450 Ry.

## 2. The evolution of free energy surface (FES)

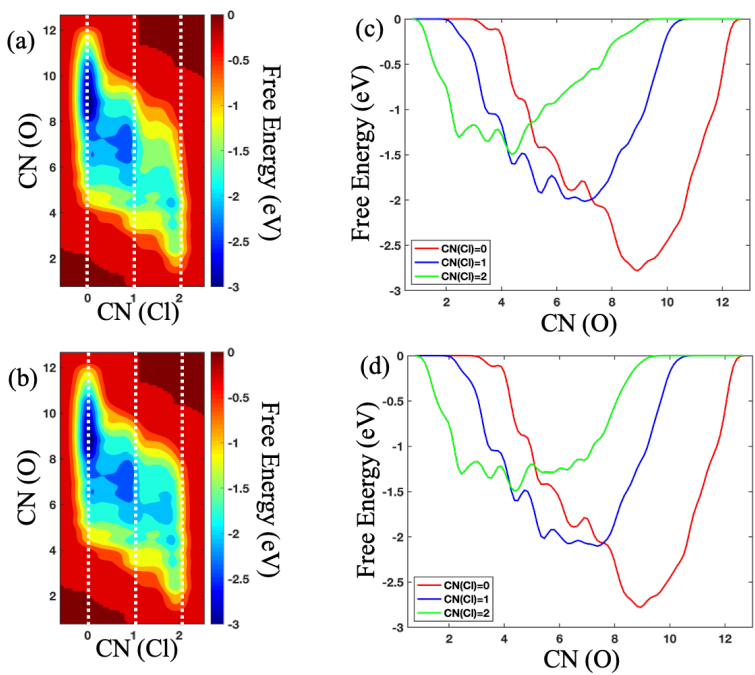

Figure S2. FES contour plots of $CaCl_2$ aqueous solution as a function of CN of Cl (*x*-axis) and O (*y*-axis) around the $Ca^{2+}$ ion from a (a) 90 ps and (b) 95 ps AIMD metadynamics simulations. (c-d) Free energy profile as a function of CN of O around the $Ca^{2+}$ ion with 0, 1, and 2 CN of $Cl^-$ ion, corresponding to the white dotted lines in (a-b), respectively.

The key features on the solvation structures (e.g., FHS, CIP, and BIP) in the FES maintains from 90 ps to 100 ps in the AIMD metadynamics simulation.

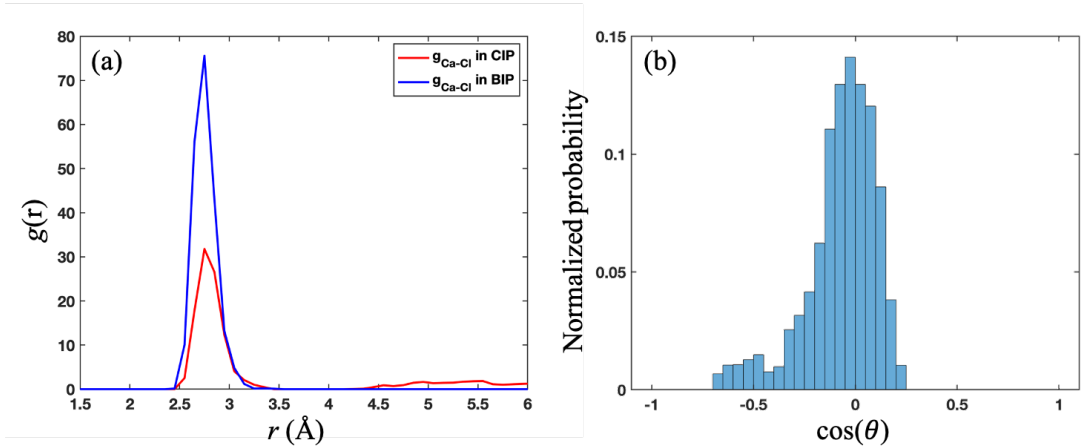

Figure S3. (a) Radial distribution function from Ca to Cl in CIP and BIP system; (b) Normalized probability of the angle distribution of ∠ClCaCl in the BIP system.

## 3. Validation on MD simulations with deep learning potential

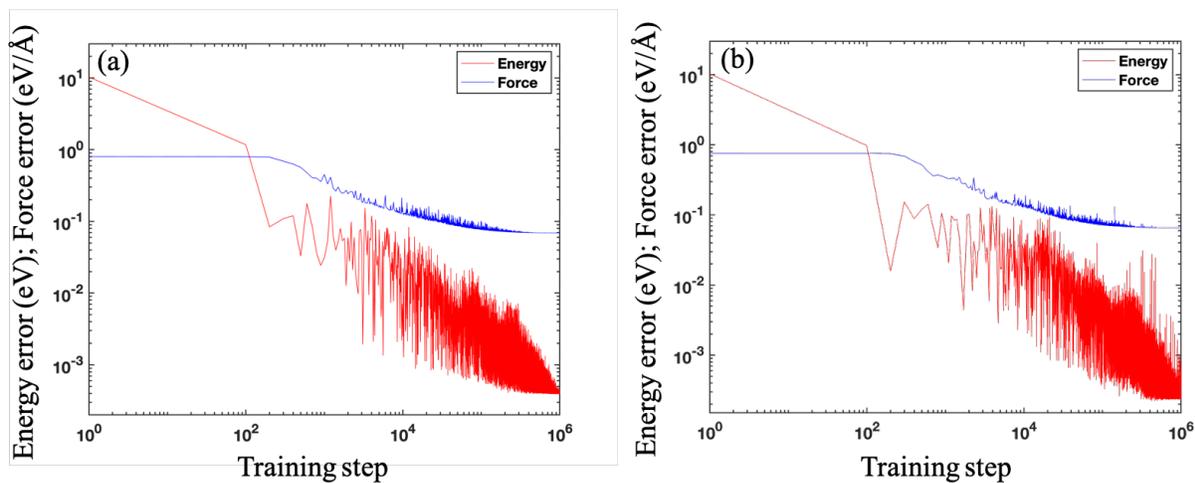

Figure S4. Learning curves of (a) bulk water and (b) $CaCl_2$ aqueous solution system. The root mean square energy and force testing errors are presented against the training step.

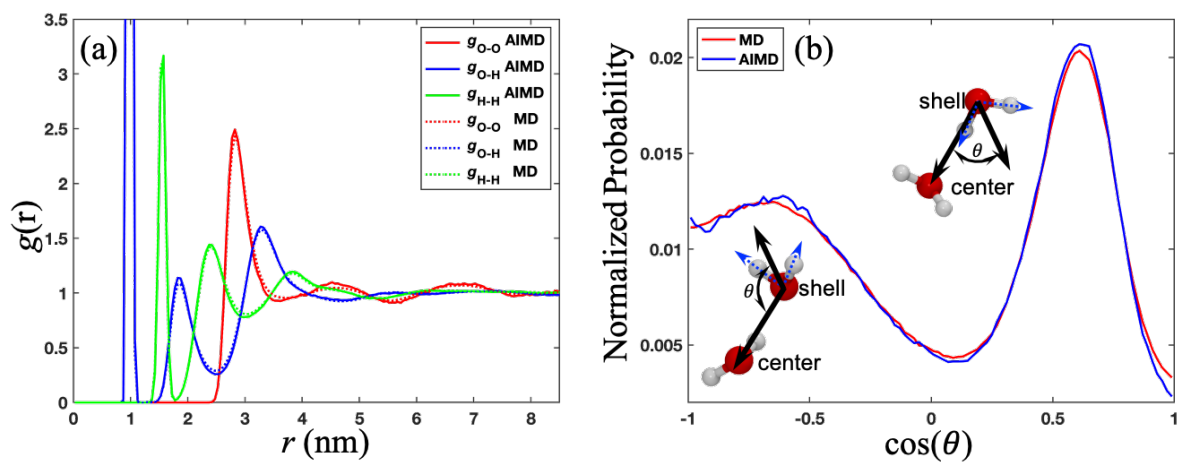

Figure S5. (a) RDF in bulk water system from AIMD and machine learning MD simulations. (b) Distribution of angle formed by $\overrightarrow{OO}$ and bisector of the neighboring water molecules in the first solvation shell from AIMD and machine learning MD simulations.

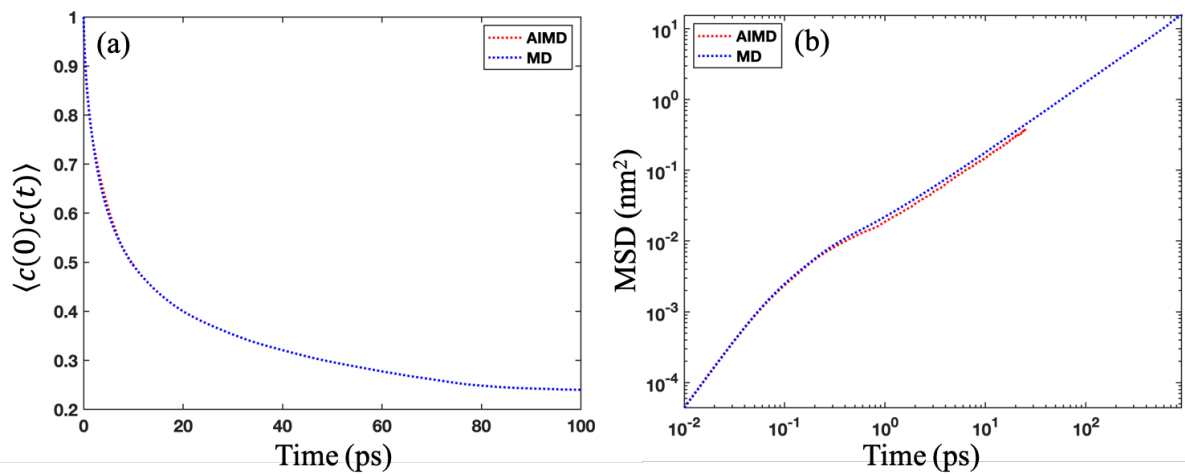

Figure S6. (a) Residence correlation function for water molecule and its neighboring water molecules and (b) mean square displacement of water molecule in bulk water system calculated from AIMD and machine learning MD simulations.

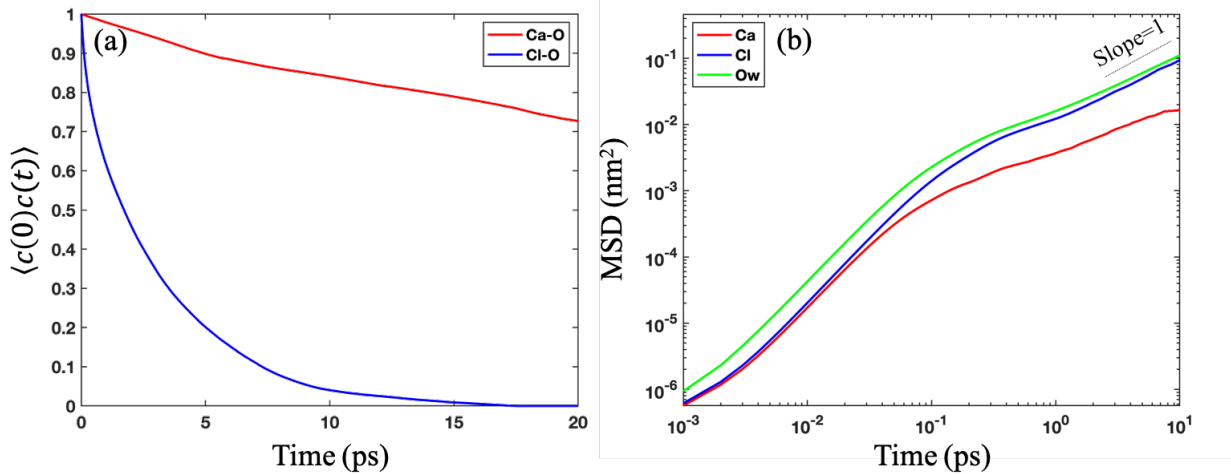

Figure S7. (a) Residence correlation function for water in the first solvation shell around $Ca^{2+}$ and $Cl^-$ ion in FHS case calculated from a 50 ps AIMD trajectory. (b) MSD of $Ca^{2+}$, $Cl^-$, and water in FHS case calculated from a 50 ps AIMD trajectory.

In this section, we provide a more extensive comparison between the results obtained from AIMD and MD with deep learning potentials. These comparisons are performed in the bulk water system with 64 water molecules and the FHS system described in the main text. The deep learning potential for the bulk water system is trained following the prior work,[1] and the learning curves are shown in Figure S5.

*Structure properties in bulk water:* From Figure S5, we can see the RDF ($g_{O-O}$, $g_{O-H}$, and $g_{H-H}$) and angular distribution function in bulk water system obtained from 20 ps AIMD and MD simulation with a deep learning potential match well. The angular distribution in the pure water exhibits two peaks, one positive and the other negative, corresponding to water molecules acting as either hydrogen bond donors or acceptors to the central water molecule, respectively.[2]

*Dynamic properties in bulk water:* From Figure S6, we can see the residence correlation function and mean square displacement (MSD) of water molecules in the bulk water system from 20 ps AIMD and MD simulation with a deep learning potential match well. The diffusion coefficients of water fitted from MSD is $2.39 \times 10^{-9}$ m$^2$/s and $2.80 \times 10^{-9}$ m$^2$/s in AIMD and MD at 303 K, respectively. Compared to the experimental data ($2.59 \times 10^{-9}$ m$^2$/s)[3], the difference is less than 10%.

*Dynamic properties in FHS system:* Figure S7 showed the residence correlation function and MSD in the FHS system obtained from 50 ps AIMD simulation, which basically matched with those calculated in the MD simulations with the deep learning potential shown in Figures 4 and 5 in the main text except the MSD of $Ca^{2+}$ ion. The diffusion coefficient of $Cl^-$ ion and water molecules fitted from MSD curves is $1.56 \times 10^{-9}$ m$^2$/s and $1.66 \times 10^{-9}$ m$^2$/s, respectively, which is similar to those calculated from MD ($1.87 \times 10^9$ and $2.28 \times 10^9$ m$^2$/s) and measured from experiments ($1.59 \times 10^9$ m$^2$/s and $1.89 \times 10^9$ m$^2$/s)[3]. The dynamics of the $Ca^{2+}$ ion is sluggish, and the MSD of $Ca^{2+}$ hasn't reached the diffusive regime (slope = 1) in the AIMD simulation. Therefore, we cannot obtain a reliable diffusion coefficient. This also explains the limitation of AIMD on the calculation of dynamics in some sluggish systems.